\documentclass[aps,pra,twocolumn,amsmath,amssymb,superscriptaddress,showpacs]{revtex4-1}

\usepackage{amsfonts}
\usepackage{amsmath}
\usepackage{amssymb}
\usepackage{amsthm}
\usepackage{fontenc}
\usepackage{graphicx}
\usepackage{xcolor}
\usepackage{textcomp}
\usepackage{epstopdf}
\usepackage{braket}
\usepackage{mathtools}
\usepackage{amsmath}
\usepackage{dcolumn}
\usepackage{multirow}
\usepackage{units}
\usepackage{array}

\begin{document}

\title{Corner and Edge States in Topological Sierpinski Carpet Systems}

\author{L. L. Lage}
\affiliation{Instituto de F\'isica, Universidade Federal Fluminense, Niter\'oi, Av. Litor\^{a}nea sn 24210-340, RJ-Brazil}

\author{N. C. Rappe}
\affiliation{Instituto de F\'isica, Universidade Federal Fluminense, Niter\'oi, Av. Litor\^{a}nea sn 24210-340, RJ-Brazil}

\author{A. Latgé}
\affiliation{Instituto de F\'isica, Universidade Federal Fluminense, Niter\'oi, Av. Litor\^{a}nea sn 24210-340, RJ-Brazil}

\date{\today}

\begin{abstract}

Fractal lattices, with their self-similar and intricate structures, offer potential platforms for engineering physical properties on the nanoscale and also for realizing and manipulating high order topological insulator  states in novel ways. Here we present a theoretical study on localized corner and edge states, emerging from topological phases in Sierpinski Carpet within a $\pi$-flux regime. A topological phase diagram is presented correlating the quadrupole moment with different hopping parameters. Particular localized states are identified following spatial signatures in distinct fractal generations. The specific geometry and scaling properties of the fractal systems can guide the supported topological states types and their associated functionalities. A conductive device is proposed by coupling identical Sierpinski Carpet units providing transport response through projected edge states which carry on the details of the system's topology. \textcolor{black}{Our findings suggest that fractal lattices may also work as  alternative routes to tune energy channels in different devices.}

\end{abstract}

\maketitle

\section{Introduction}
\textcolor{black}{Fractal systems are abundant in nature, characterized by their self-similar patterns and complex repeating structures at fractional dimensions.} While they have been extensively researched in the past, primarily concerning the self-similarity properties \cite{Kim2006}, they are presently being studied within the framework of topological materials. Scanning tunneling microscopy \cite{Kempkes2019} and chemical routes \cite{Shang2015, mo2019surface, 10.1093/nsr/nwad088} have successfully produced fractal samples at the molecular scale. Furthermore, several proposals have reported topological insulator features in fractal families \cite{Biesenthal2022, Junkai2022, Zheng2022observation}.

Topological regimes can be achieved by external perturbations and modulations of the electronic properties. Models such as Haldane \cite{PhysRevLett.61.2015}, Kane-Mele \cite{kane2005}, and others have been used in graphene-like lattices to explore the emergence of topological phases by tuning bulk-boundary correspondences in electronic structures. The modern polarization theory \cite{resta,vanderbuilt1,vanderbuilt2} has changed the understanding of electronic localization within crystals, especially regarding surface/edge and bulk charge accumulation. In this context, the Bernevig-Benalcazar-Hughes (BBH) model describes electrons on a square lattice with $\pi$-flux per plaquette and dimerized hoppings along the x and y directions \cite{Benalcazar_2017,2Benalcazar_2017}. This leads to quadrupole (2D) and octupole (3D) polarization, resulting in fractional charges at the corners (2D) and hinges (3D). These discoveries have paved the way for a new class of topological insulators known as higher-order topological insulators (HOTIs), where these properties are linked to the system's dimensionality \cite{Benalcazar_2017,2Benalcazar_2017,hua2023,hoti}. The validity of these systems has been confirmed through theoretical studies \cite{khalaf2021boundary,Arouca2020,Peterson2020} and experimental setups such as acoustic lattices \cite{Xue2020,ni2020demonstration,PhysRevResearch.5.023189}, electrical circuits, and metamaterials \cite{lv2021realization,bao2019,Peterson2020ObservationOT}.

When considering a fractal system and its fractional dimension, a particular issue that arises is how the dimension impacts HOTI classification and the fractional charge features in these new geometries. A special signature of HOTI in these fractal systems is the existence of inner corner states arising simultaneously with the outer corner ones \cite{pai, PhysRevB.105.L201301, PhysRevB161116}.

Other topological characterizations, and Anderson disorder, were also studied in such fractal systems \cite{PhysRevB.98.205116,Chin2024}. Highly localized electronic states with strictly vanishing amplitude outside a finite lattice region and well-defined in terms of energy are known as compact localized states (CLS). In periodic lattices, these states are sometimes related to flat bands, a non-dispersive topological state in the energy structure \cite{chen2023decoding,PhysRevB.95.115135}. This compact localization results from an intricate interplay between lattice geometry and electronic interaction, causing destructive interference and contributing to the emergence of confined states \cite{Li2018,Lim2020,Lazarides2019}. Such features are also manifested in finite systems and have been reported in fractal regimes defined by triangle Sierpinski lattices \cite{Cristiane2}.

Transport properties in periodic systems with fractal dimensions have been previously investigated in non-topological regimes, particularly in quasi-one-dimensional and molecular fractal chains \cite{PhysRevResearch.2.013044, PhysRevB.93.115428, PhysRevResearch.3.043103,D2CP02426H, Frontier2023}, based on the experimental synthesis of such lattices \cite{exp1, exp10}. Various underlying lattice structures have been employed to model nanofabricated Sierpinski Carpets (SC), reported in conductance fluctuation analysis \cite{Han2019} and in functionalized fractal proposals \cite{PhysRevB.105.205433}. Here we propose the formation of molecular fractal chains consisting of topological SC repetitions. Transport conductance was analyzed in specific topological phases of periodic SC. Our findings reveal the presence of conductive channels within the dipolar and quadrupolar phases of the fractal chains, underscoring the robustness of this topological structure.   

In the current work, we provide a comprehensive analysis of Sierpinski Carpet (SC) fractal geometries, emphasizing the localized characteristics of the electronic state. Through a combined spectral charge analysis, we shed light on the system's topological properties. By investigating the energy spectrum and charge distribution across different lattice configurations defined by hopping energy parameters, we construct a detailed topological phase diagram. This analysis uncovers a topological transition between quadrupolar, dipolar, and trivial phases, each distinguished by unique and new charge densities at the corner/anti-corner, edge, and bulk-like states, all adhering to $C_4$ symmetry.

\section{Theoretical Model}

To describe the square flakes we use a spinless single-orbital  tight-binding (TB) Hamiltonian, 

\begin{equation}
H=\sum_{i}\varepsilon_{i}c_{i}^{\dagger}c_{i}+\sum_ {\left\langle ij\right\rangle} t_{ij}c_{i}^{\dagger}c_{j}+h.c.,
\end{equation}
with $\varepsilon_i$ being the on-site energy, $c^{\dagger}_i$ ($c_i$) the creation (annihilation) operator of an electron at site $i$, and $t_{i,j}$ the hopping energies for nearest neighbors sites, that are separated as intra- and intercell hopping considering as an unit cell a single 4-fold square \cite{Benalcazar_2017, Arouca2020}. For the square lattices discussed here, they are  written as $\gamma_{x(y)}$  and $\lambda_{x(y)}$ parameters, respectively. The electronic properties of the studied SC systems are derived via eigenfunctions $\psi_n$ and eigenenergies $\epsilon_n$ calculations, with $n$ being the number of atomic sites. A localization parameter $\mathcal{L}_n$ is defined computing the wave function amplitude over particular $i$ sites for the $n$-th eigenenergy,
   $\mathcal{L}_n=\sum_i |\psi^{i}_n|^2$. 
A spectral charge may be defined by summing $\mathcal{L}_n$ across all occupied n-th bulk states and all i-th sites inside the unit-cell of each region of the space \cite{Li2022},

\begin{equation}
Q=\sum_{n}^{bulk} \mathcal{L}_n\,\,.\label{charge}
\end{equation}

\textcolor{black}{For the BBH model, the robustness of the corner charges can be verified by calculating the quadrupole moment $\mathcal{Q}^{x y}$  which is the topological invariant that is quantized by the value of the corner charges for a given HOTI regime in the parameters. It is expressed as \cite{expression,PhysRevB.100.245134,PhysRevB.100.245135},}

\begin{equation}
    \mathcal{Q}^{xy}=\frac{1}{2\pi}\text{Im}{ \ log\bigg[det(U^{\dagger}qU)\sqrt{det(q^{\dagger})}\bigg],}\label{eq3}
\end{equation}
\textcolor{black}{with U being a matrix with half filling columnwise eigenvectors and $q=$exp$( 2i\pi \textbf{xy} $/$ L_x L_y )$ with $\textbf{x(y)}$ being the position operator and $L_{x(y)}$ the $x(y)$ coordinate length.}

For periodic systems, transport properties are derived following the Landau formalism \cite{Datta1995}, in which the conductance of the system is given by G(E)=$2e^2/h\mathcal{T}(E)$, with the transmission $\mathcal{T}(E)$  written  in terms of the Green function of the scattering center and leads, and is given by,

\begin{equation}
{\mathcal{T}(E)}=\operatorname{Tr}\left[\boldsymbol\Gamma^{L} \boldsymbol{G}^{r}_{c} \boldsymbol\Gamma^{R} \boldsymbol{G}^{a}_{c} \right]\,\,,
\end{equation}

\noindent with
 
\begin{equation}
G^{a,r}_{c}(E)=\Big[\omega - {H_c} - \Sigma^{a,r}_L(E) - \Sigma^{a,r}_R(E)\Big]^{-1}\,\,,
\end{equation}
where $\omega = E \pm i\eta $, $\eta$ being an infinitesimal energy value and  ${H}_{c} $ the Hamiltonian of the central part. $\Sigma^{a,r}_{L,R} (E)$ correspond to left and right self-energies, given by the  related surface Green functions, from which the coupling matrices are obtained via $\Gamma^{L,R} (E)= i (\Sigma_{L,R}^{r} (E) - \Sigma_{L,R}^{a}(E))$. 
\textcolor{black} {Here the leads are composed by the same system as the scattering center, i.e., Sierpinski carpet repetitions.} 
\textcolor{black}{Numerically,  to derive the electronic properties described within the real-space tight binding, standard softwares were used, such as Mathematica and Fortran, but own codes were developed in all calculation processes. The Green's function of the semi-infinite leads are obtained numerically by a decimation method \cite{Latge}.}

\section{Results and Discussions}

Electronic properties and quantum transport in Sierpinski Carpet fractals are calculated in the following. In the topological case we consider a priori $\gamma_x$ and $\gamma_y$ as variable parameters, assume $\lambda=\lambda_x=\lambda_y$ constant equals to 1.0, and all the energies are given in terms of $\lambda$.  Before discussing the topological SC lattice  let us point out some peculiarities of the fractal system related to the choice used for mapping the internal sites, i.e., the partition net adopted.

\subsection{Non-Topological Sierpinski Carpet}

\begin{figure}[!h]
\centering
\includegraphics[width=8cm]{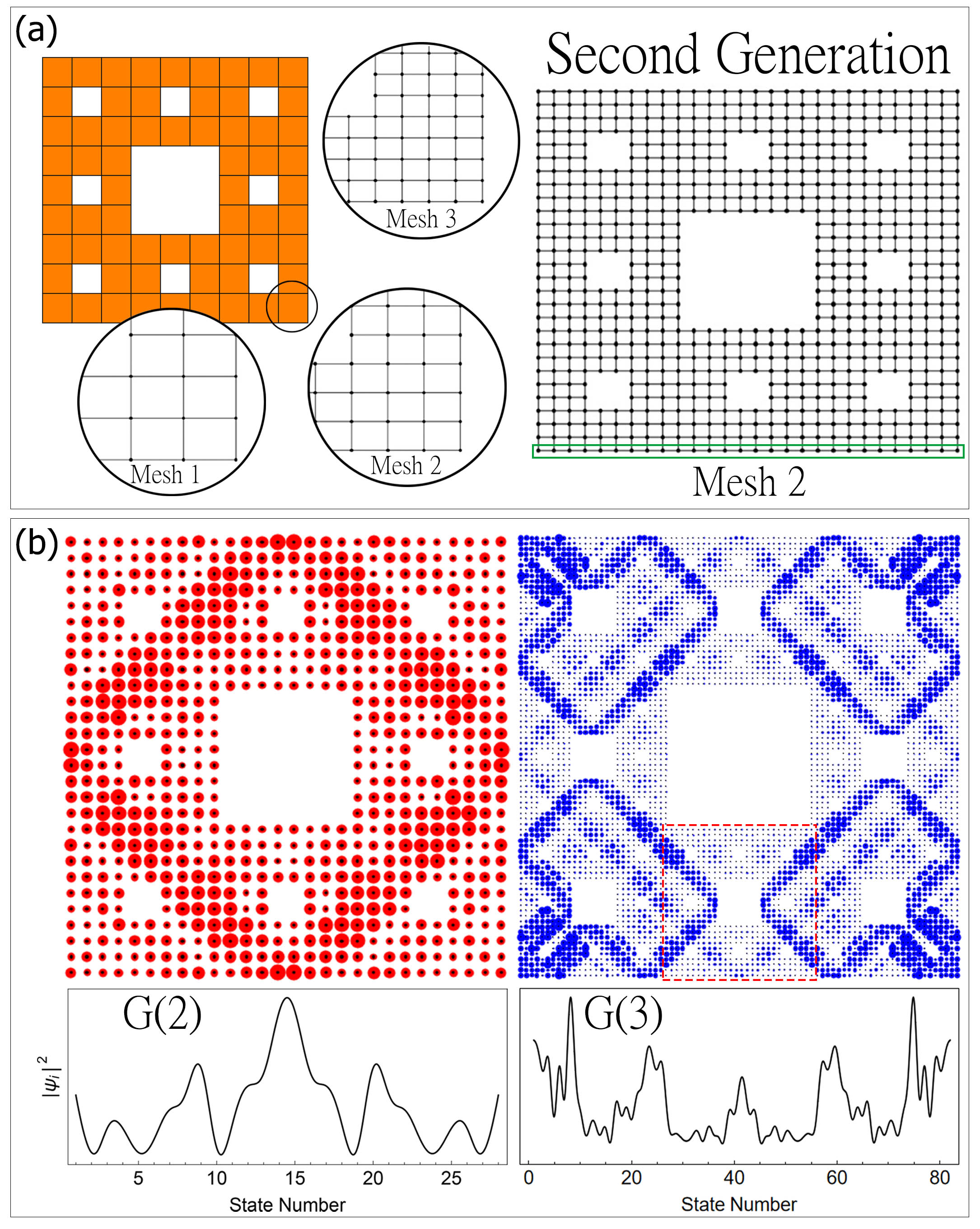}
\caption{ (a) Schematic view of a second order SC. An unit cell is marked by a small circle in the low-right corner. The larger circles illustrate pieces of the SC with increasing number of lattice partitions [Mesh 1, 2 and 3]. At right a G(2) Sierpinski carpet described using Mesh 2.(b) LDOS at the Fermi energy $E=0$, for G(2) and G(3) SCs, using Mesh 2.  At the bottom part corresponding amplitude density at the edges exhibiting a self-similar pattern.}
\label{nontoposc}
\end{figure}

The SC is a fractal structure constructed by creating holes inside a solid square region via a fixed protocol.
As displayed in Fig. \ref{nontoposc}(a) the first generation contains 8 solid tiles, with an internal hole between them, derived from the zero-order square by dividing each length by a factor of 3. The fractal dimension or Hausdorff dimension in this case is calculated as $d_1=log(8)/log(3)\approx1.89$. For the second generation we have 64 tiles and edges divided by 9, yielding also $d_2=log(64)/log(9)\approx1.89$. Therefore after the n-th iteration, the fractal dimension is $d_n=log(8^n)/log(3^n)\approx1.89$ independently of the desired \textcolor{black}{n-th} generation. 

Apart from the mathematical aspects, the same principles apply when constructing physical fractal systems. From a tight-binding perspective, investigating the electronic properties of SC structures requires describing these systems in terms of lattice partitions. It is important to note that various geometric arrangements can be employed for such calculations \cite{PhysRevB.93.115428,PhysRevB.105.205433}. A natural starting point is the square lattice, with different mesh point densities highlighted through zoomed-in views (circles) in Fig.\ref{nontoposc}(a), referred to as Mesh 1, 2, and 3. The choice of mesh and the algorithm used to generate these systems are crucial to maintaining the symmetries inherent to the fractal structure.

In Appendix A, we provide a detailed discussion on mesh point density and its implications for electronic density distribution and state degeneracy. For example, Fig. \ref{nontoposc}(b) shows the local density of states (LDOS) at the Fermi level for the second [G(2)-red] and third [G(3)-blue] SC generations using Mesh 2. A self-similar pattern emerges between the second and third generations when comparing the projected wave function, $|\psi|^2$, along the corresponding edge sites, as highlighted by the green rectangle in Fig. \ref{nontoposc}(a). The projected states in the middle of this edge exhibit self-similarity across generations. In the subsequent sections, we adopt Mesh 2 numerically to explore the electronic properties of the topological SC.

\subsection{Topological Sierpinski Carpet}

As discussed, HOTI phenomena was predicted to exists as well in fractional dimensions. 
The existence of compact localized states were  explored in confined systems \cite{Cristiane2}, in which, the CLS are separated from the bulk states. Here, they have been seen for dipolar and quadrupolar phases. \textcolor{black}{We stress that in the following discussions, the colors blue, green, orange and gray are associated with each phase depicted in the Diagram. Few exceptions are explicitly mentioned in the text.}      

\begin{figure}[!h]
\centering
\includegraphics[width=8.6cm]{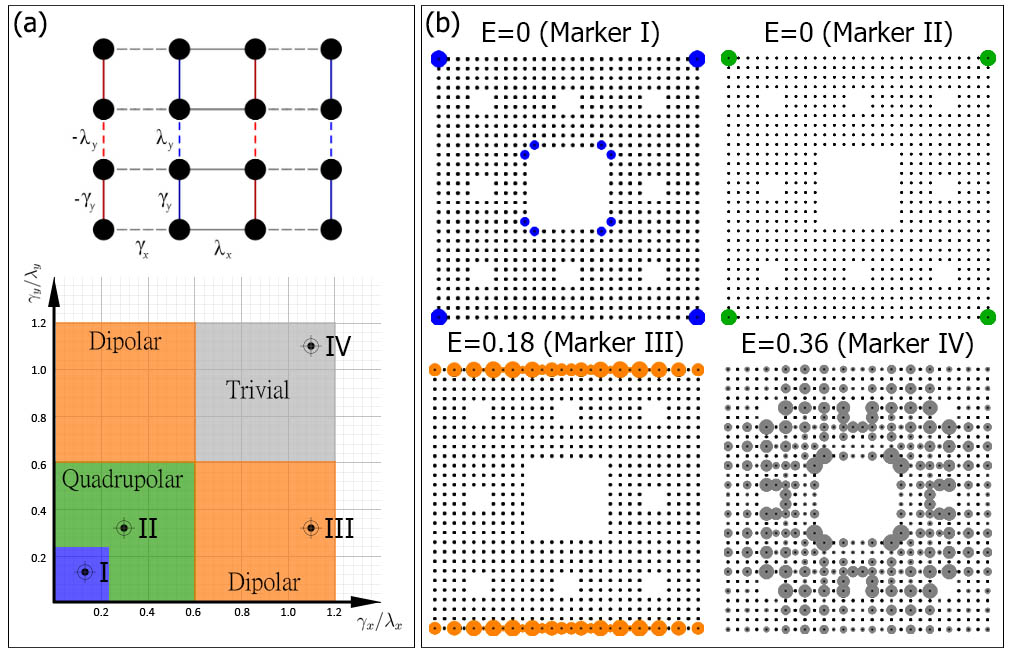}
\caption{(a) $\pi$-flux hopping scheme and topological diagram. (b) LDOS at E=0 for Marker I ($\gamma_{x(y)}=0.17$) and II ($\gamma_{x(y)}=0.37$), and at the first nearest eigenenergy from the Fermi level, in the cases of Marker III ($\gamma_{x}=1.10$ and $\gamma_{y}=0.37$) and IV ($\gamma_{x(y)}=1.10$).}
\label{diagrama}
\end{figure}

\begin{figure*}
    \centering
    \includegraphics[width=16cm]
   {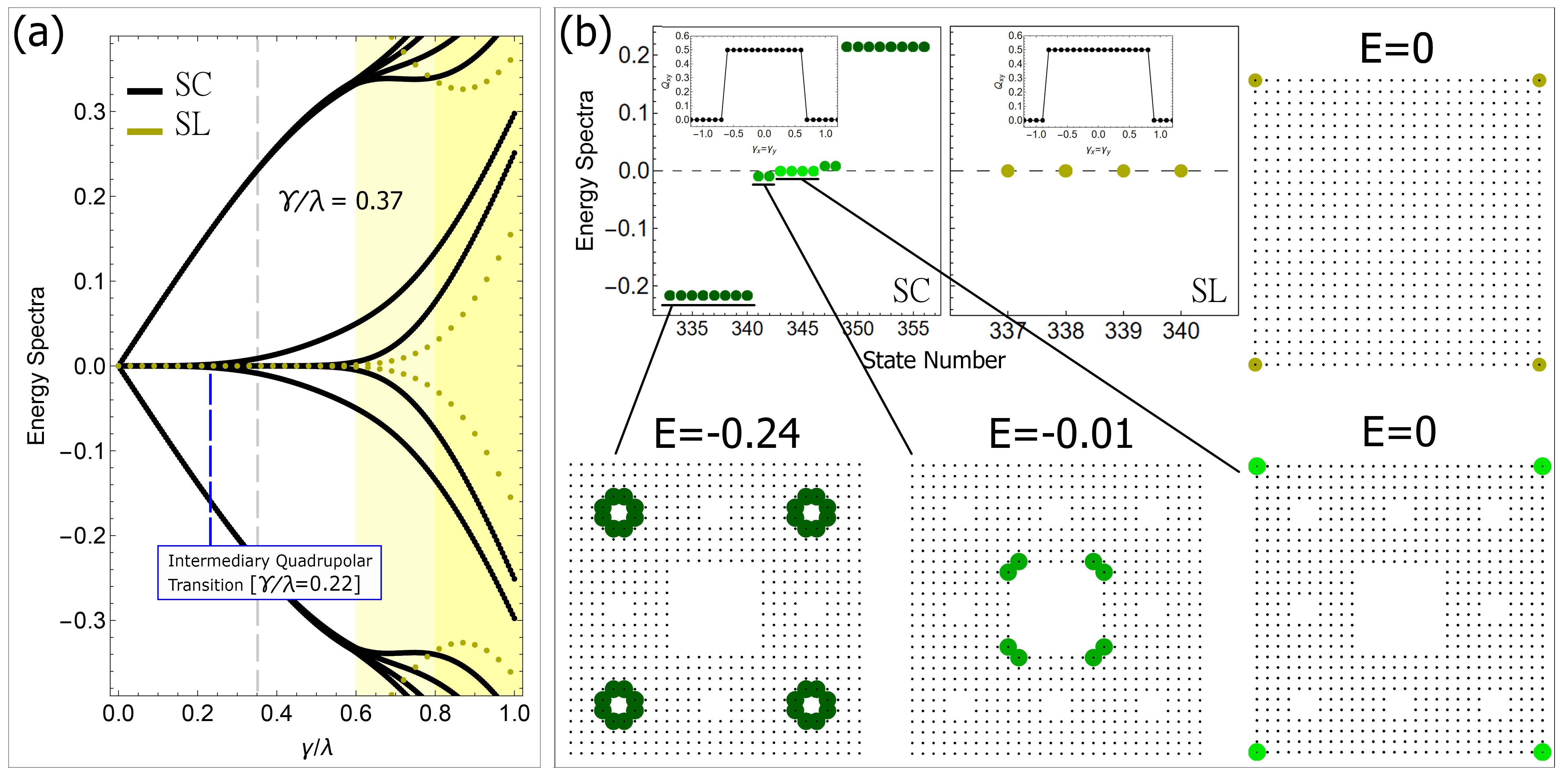}
\caption{(a) Energy spectra plot for  $\gamma/\lambda$ in both topological Sierpinski carpet (SC) and square lattice (SL) with lighter and darker yellow sections denoting the topological-trivial transition, respectively, for SC and SL, with equivalent number of sites. (b) Zoom in the Energy spectra for the dashed gray line at $\gamma/\lambda=0.37$ and LDOS in shadows of green Inset: Quadrupole moment versus the hopping energies $\gamma_{x(y)}$, highlighting the quadrupolar to trivial transition for SC and SL. The circles in LDOS are proportional to the normalized square modulus $|\psi_i|^2$ in each case.}
\label{energyvsga}
\end{figure*}

The hopping energy distribution emerging from the adopted  $\pi$-flux model is illustrated in Fig. \ref{diagrama}(a); blue and red vertical lines have positive and negative values, respectively. By changing the $\gamma/\lambda$ ratio in both x- and y-directions, it is possible to identify  topological and trivial phases as displayed in the phase diagram at the bottom part of  Fig. \ref{diagrama}(a), following the colored pattern domains. The topological-trivial phase transition is determined using the quadrupole moment Eq. \ref{eq3} which gives a quantized $\mathcal{Q}^{xy}=0.5$ value for $\gamma_{x(y)}=0.6$.
\textcolor{black}{Four particular points, identified as marker positions I, II (Quadrupolar) III (Dipolar) and IV (Trivial) are highlighted inside each respective diagram region. The corresponding LDOS is computed at the Fermi level for Markers I and II, and for the eigenenergies closest to the Fermi level for Markers III and IV, as depicted in Fig. \ref{diagrama}(b). These calculations reveal the spatially dependent nature of the wave functions for each set of parameter ratios, showing localization at the corners, edges, and across the center-symmetric region of the SC}.  The LDOS are normalized by the wave function local maximum value and show two distinct charge symmetries $C_4$ and $C_2$ whether the system is in a quadrupolar/trivial or dipolar phase, respectively. The size and color of each filled circle represent the wave amplitude and phase in the diagram of Fig. \ref{diagrama}(a), respectively. 
Differently from the square lattice results \cite{Arouca2020,khalaf2021boundary}, there is an intermediary transition inside the quadrupolar phase 
where it is possible to distinguish between pure corner states, in green, and mixed outer corner with inner anti-corner localized states, in blue. 
Moreover, the topological-trivial transition happens for smaller $\gamma$ values for the fractal lattice in comparison with the square lattice. To clearly identify the topological phase state in each scenario, this color scheme will be used in the analysis of the electronic property findings. In conjunction with the LDOS, the energy spectra is fundamental to clarify this transition, as it is discussed in the following.

 Fig. \ref{energyvsga}(a) shows the energy spectra of a G(2)-SC as a function of the $\gamma/\lambda$ ratio for the states close to the Fermi level.  For comparison, the results corresponding to a square lattice (SL) of the near size are also displayed (dotted golden curves).  
It is worth noting that the black lines (SC) are degenerated at $E=0$, splitting into three energy values beyond the intermediary transition point at $\gamma/\lambda=0.22$, highlighted by the blue dashed line as a quadrupolar transition. \textcolor{black}{The quadrupolar-trivial transition is highlighted in the energy spectra with lighter ($\gamma=0.6$) and darker ($\gamma=0.8$) yellow shaded regions, respectively, for SC and SL flakes}. In Fig. \ref{energyvsga}(b) the quadrupole moment is shown in the inset of the zoom at the energy spectra for near sized SC(SL) flakes, being quantized as $\mathcal{Q}^{xy}=0.5$, for $\gamma_{x(y)}$ up to 0.6(0.8) \textcolor{black}{giving raise to the quadrupolar-trivial phase transition in such systems}. Choosing $\gamma/\lambda=0.37$, it is possible to identify such transition, as shown in Fig. \ref{energyvsga}(b) for both SC and SL; the estimated charge densities for this choice are shown with different green shades for the SC and with golden circles for the SL in the LDOS. The states for the SC in this phase, localized at the green region (Marker II) in Fig. \ref{diagrama}(a) are divided by pure corner states spatially distributed at the outer corners ($E=0$), inner anti-corner states at the G(N-1) hole ($E=0.01$) and inner anti-corner charges localized in four of the 8 G(N) holes ($E=0.24$), with 4, 2, and 8 fold degeneracy, respectively. 
We adopt a particular terminology for the SC holes to refer to different levels of fractal hierarchy, i.e., G(N) and G(N-1) holes are used to define the new holes emergent from the N-generation in contrast to the previous holes already present in the preceding generation, respectively. 
All such charge distributions preserves the center-symmetry in relation to the G(N-1) hole, following the $C_4$ symmetry. Interestingly, even in the presence of a symmetry break on the regular geometrical arrangement of the fractal SC the center-symmetry of the electronic distributions around the holes is preserved, as it is shown in Appendix B. Also discussed in the same Appendix is the case of the electronic property outcomes including the spatial  localization of the emerging  states for a G(3)-SC.

\begin{figure}[!h]
    \centering
    \includegraphics[width=8.6cm]{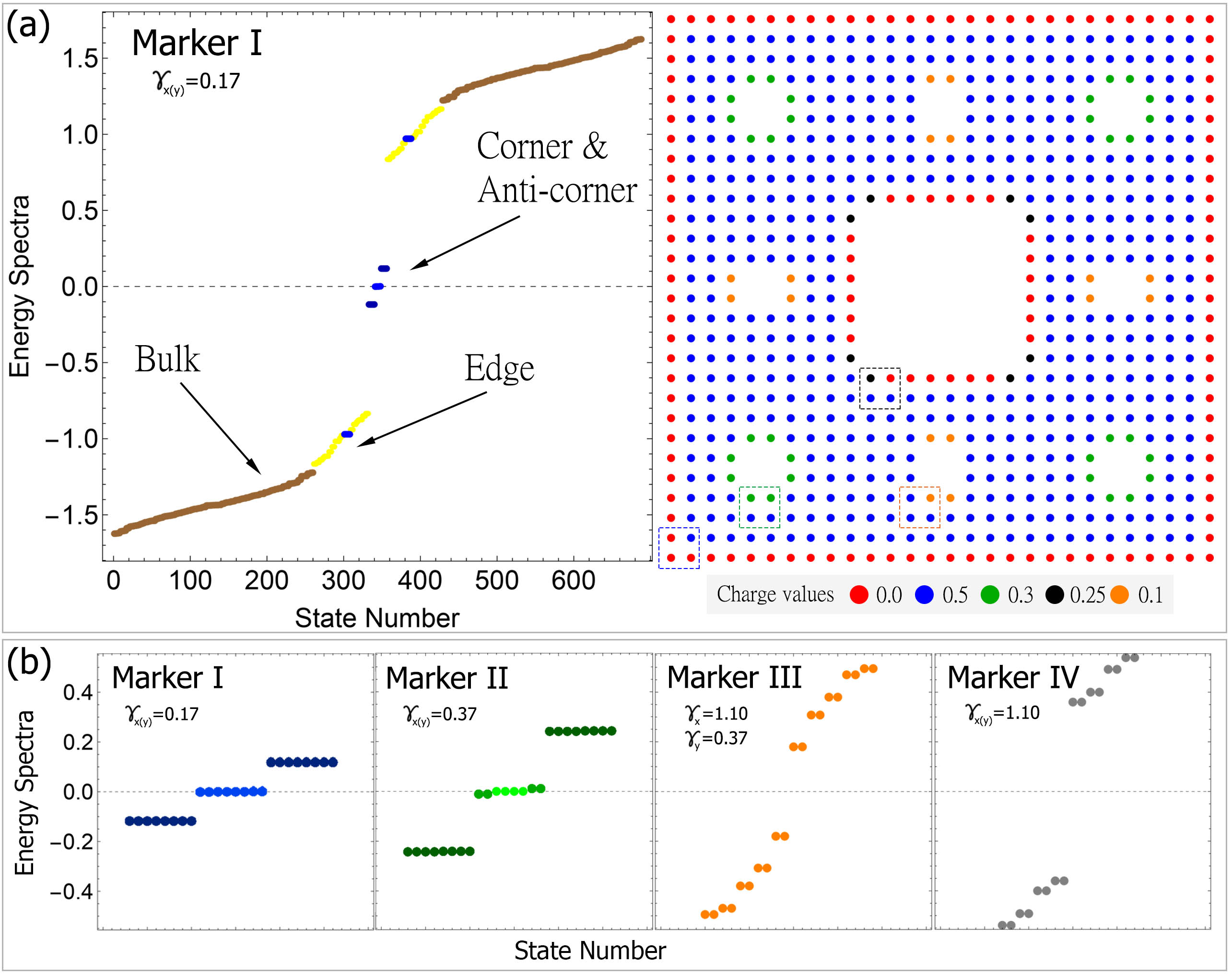}
   \caption{(a) Left: G(2)-SC Energy spectra, with corner, edge and bulk states highlighted in blue, yellow and brown, respectively. Right: Spectral charge $Q$ calculated at each SC site. In both calculations $\gamma_{x(y)}=0.17$. (b) Zoom view near the Fermi level of further parameter combinations; the hopping $\lambda_{x(y)}=1.0$ in all calculations.}
   \label{energyspectra}
\end{figure}

A general topological phase characterization of a G(2)-SC is presented in Fig. \ref{energyspectra} via the energy spectra information and spectral charge value $Q$. The energy spectra displayed in panel (a) is derived for $\gamma_{x(y)}=0.17$ and $\lambda_{x(y)}=1.0$, corresponding to Marker I position in Fig. \ref{diagrama}(a). \textcolor{black}{The spatial distributions of the eigenstates at the bulk, corners, and edges sites are identified by brown, blue, and yellow colors, respectively}. Within the quadrupolar phase region of the diagram,  Marker I and Marker II  (blue and green area in the diagram phase exhibited in Fig. \ref{diagrama}), where the system is a HOTI, with the corner, edge and bulk states coexisting in the complete energy spectrum. Mid-gap states  within the bulk and edges boundary, shown in Fig. \ref{energyspectra}(b), are typical responses in this regime. Otherwise, in the dipolar and trivial phase, the system exhibits exclusively edge- and bulk-like states distribution. Furthermore, comparing the zoom in the energy spectra next to the Fermi level for both quadrupolar phases, Marker I and II, in blue and green, respectively, it is possible to see in the case of Marker I that the $E=0$ state is 8x degenerated. For Marker II the state degeneracy is 4-fold, as shown in the second panel of Fig. \ref{energyspectra}(b). This difference is related with the mentioned transition that takes the inner anti-corner states apart from the outer corner ones in the quadrupolar phase. Moreover, Fig. \ref{energyspectra}(b) exhibits also the energy spectra for a dipolar and a trivial phase (Markers III and IV), marked by the presence of energy gaps \textcolor{black}{due to the break of $C_4$ symmetry}.

The spectral charge is calculated by summing over the bulk energy states [brown symbols in Fig. \ref{energyspectra}(a)], as stated in Eq. \ref{charge}.
 We can distinguish the charge $Q_i$ calculated at the i-th site inside the unit cell of each region, being, outer corners at the external edges and anti-corners at both SC G(N) and G(N-1) holes. \textcolor{black}{For a regular square lattice at HOTI phase, the computed spectral charges at the corners yield $Q_{cor}=0.5$ \cite{2Benalcazar_2017,Benalcazar_2017}. The spectral charges for individual SC sites are shown in the right panel of Fig. \ref{energyspectra}(a).} First, by computing the charges of the outer corners inside the dashed unit cell at the extreme left-bottom position, we find $Q_{cor}=0.5$, as expected for a quadrupolar HOTI phase. The anti-corner charges for the black sites in the central G(1) hole differ from those for the green and orange sites in the G(2) holes. By summing the contributions of the corresponding unit cell sites, marked in the colored map, we obtain for the anti-corners $Q_{G(1)}=1.25$ and $Q_{G(2)}=1.6$, both being fractional values \textcolor{black}{which is a new feature in HOTI SC}. For the non-topological bulk regions of the square lattice (SL) an integer value of the charges is expected, as obtained here for the SC as \textcolor{black}{$Q_{edge}=1.0$ and $Q_{bulk}=2.0$. Out of the quadrupolar phase the distinction between the bulk and corner/edge states are mixed leading to indistinguishable spectral charge situations.}
 We emphasize here that the particular lattice mesh plays a special role in defining the spectral charge values. However, although other lattice partitions may lead to different spectral charge distribution, the overall calculations predicts HOTI phases with fractional charges for such corner and anti-corner states \cite{Junkai2022,Zheng2022observation}. 

Despite the bulk states (brown eigenvalues) being kept apart in energy from the corner and anti-corner states laying close to the  Fermi level, the calculation of the bulk spectral charge reveals a spatial charge 
distribution in G(2) and G(1) holes very close to the local density of states for the states at the Fermi level [see Fig. \ref{energyvsga}(b)]. This may indicate a connection between corner and bulk states in finite systems as already verified for HOTI square lattices \cite{PhysRevB161116}. 

\begin{figure}[!h]
\centering
\includegraphics[width=8.3cm]{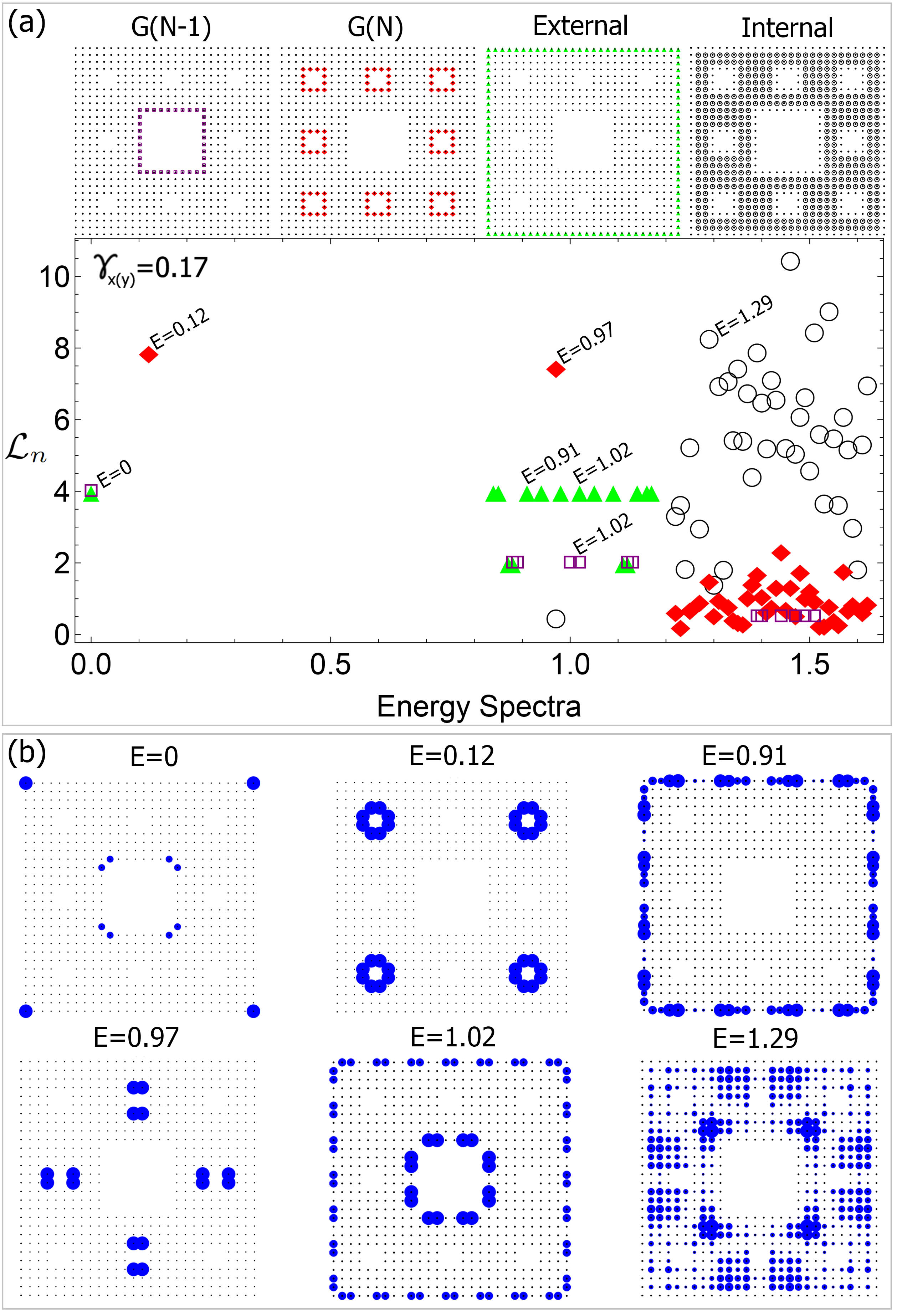}
\caption{(a) Localization number $\mathcal{L}_n$  vs energy spectra  of a G(N=2) SC, for the Mark I point in the blue region in Fig. \ref{diagrama}(a) [$\gamma_{x(y)}=0.17$ and $\lambda_{x(y)}=1.0$]. The SCs on top illustrate the segregated real space regions, defined as G(N-1=1) and G(N=2) holes, external and internal regions, with purple, red, green and black symbols, respectively. (b) LDOS at particular energies, E=0, 0.12, 0.91, 0.97, 1.02 and 1.29 marked in the $\mathcal{L}_n$ map.}
\label{locator}
\end{figure}


A complementary description of the quadrupolar phase is obtained computing the energy states weight in the spectra, as defined by the localization marker $\mathcal{L}_n$. Actually, it helps shedding light on the spatial charge distribution and wave amplitude density. The  states are classified in the inset of Fig. \ref{locator}(a), as localized at G(N-1) and G(N) holes, external, and internal regions, painted with purple, red, green, and black symbols, respectively. Since the system is electron-hole symmetric, only positive energies are shown. It is important to note that G(N) and G(N-1) hole states contain both corner and anti-corner sites, as well as edge contributions; in fact, they include all sites surrounding the SC holes.

Similarly, the external states comprise edges and outer corner sites, represented by green  triangle symbols in the figure. At $E=0$, the SC outer corner states coexist with  G(N-1) hole states (empty purple square), with $\mathcal{L}_n \approx 4$ for each case. This feature is highlighted in the LDOS for $E=0$ presented in Fig. \ref{locator}(b). Also, external sites and G(1) holes states contributes to the LDOS at $E=1.02$.  
Concerning to  G(N=2) holes (red diamond symbols) the value  $\mathcal{L}_n \approx 8$ is found for both $E=0.12$ and $E=0.97$ states. It is important to mention that such states are 8-fold degenerated as can be verified in Fig. \ref{energyspectra} emphasizing the compact localized feature of such states. Furthermore, in terms of LDOS spatial charge configuration, they are complementary, since the missing states at the G(2) holes in one instance appear in the other energy value. This is not the case of the others G(2) hole states,  happening at higher valued eigenstates together with internal states, but contributing with  small localization numbers. While internal states (empty black circles) appear almost exclusively  for $E> 1.17$, external states (green triangles) are basically presented in the range of $0.80<E<1.17$, with exception for the external corner contribution  at $E=0$. Two example of such external states, $E=0.91$ and $E=1.02$, are shown in Fig. \ref{locator}(b) reflecting a decreasing localization aspect, spread along the edges.

When the fractal symmetry is broken, the two $\mathcal{L}_n \approx 8$ states for G(N) holes collapse to the same energy value, as discussed in Appendix B. The overall picture in Fig. \ref{locator}(a) may change as we move through the phase diagram in Fig. \ref{diagrama}(a), making it more complex to distinguish the charge distribution for the energies in each region.

\subsection{Transport in Periodic Topological SC}

\begin{figure}[!h]
    \centering
    \includegraphics[width=8.5cm]{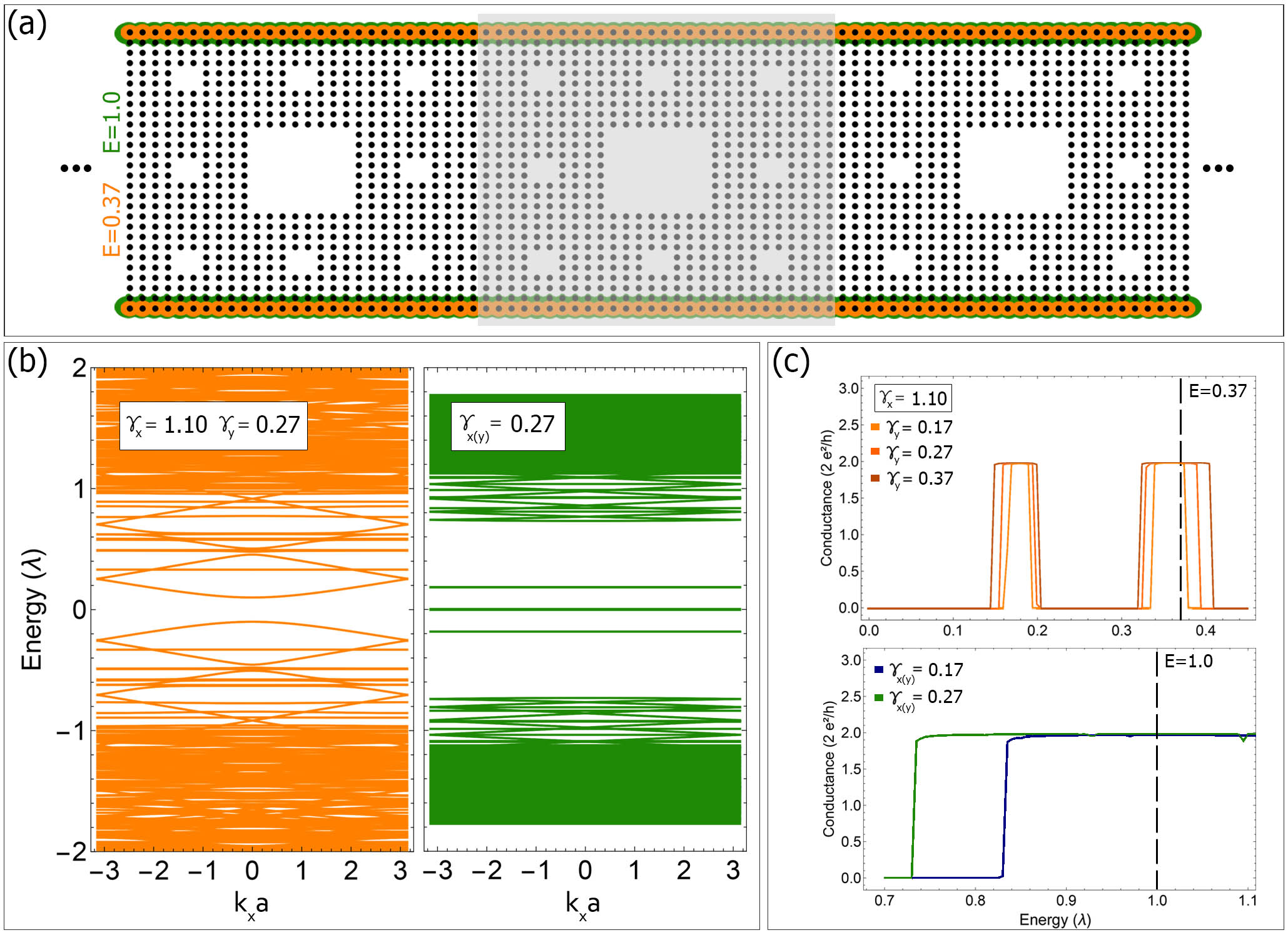}
   \caption{(a) Schematic view of a G(2)-SC chain with LDOS edge states verified at the energies $E=0.37$ and $E=1.0$ for the dipolar (orange) and quadrupolar (green) phases. (b) Band structures from left to right for $\gamma_{x}=1.10$ and $\gamma_{y}=0.27$, and for $\gamma_{x(y)}=0.27$. (c) Conductance results for both phases and different $\gamma_{x(y)}$ values; $\lambda_{x(y)}=1.0$ for both calculations}
   \label{transport}
\end{figure}

The possibility of obtaining edge states in the topological SC, as discussed in the previous sections, offers the opportunity to propose a fractal transport device. Here, we describe quasi-1D molecular chains derived by using boundary conditions in the unit cell defined as a single G(2)-SC [shaded region in Fig. \ref{transport}(a)], and investigate how the electronic states propagate in such fractal systems composed of a SC array.

The electronic structures for such devices are shown in Fig. \ref{transport}(b) for a fixed $\lambda_{x(y)}=1.0$ parameter and two examples of $\gamma_{x(y)}$ hopping. In the first case, $\gamma_x=1.1$ and $\gamma_y=0.27$ (dipolar phase), while in the second example, $\gamma_{x(y)}=0.27$ (quadrupolar phase). The single bands at and near the Fermi level in the green band structure do not contribute to transport responses. 
\textcolor{black}{In particular, the highly localized nature of the states in the energy range $-0.7<E<0.7$, for $\gamma_{x(y)}=0.27$,  results in non-dispersive electronic bands (flat bands/green curves) for the case of periodic systems.  The electronic charge distributions are located at the corners of the supercell replicas (not shown) appearing at special sites in the lateral edges, as a reminiscence feature of the corner states of the finite SCs.} On the other hand, the emergence of dispersive bands in the dipolar phase at lower energy values gives rise to available electronic channels, reflected in the plateaux in the conductance graphs (orange curves). The energy range of the conductance plateaux can be manipulated by tuning the $\gamma_{x(y)}$ values, as illustrated in Fig. \ref{transport}(c).

The quadrupolar phase exhibits a separation between corner, edge, and bulk states, as shown in Fig. \ref{energyspectra}(a), represented in finite structures as discrete energy spectra. For the periodic system, this distinction is also evident, with edge states being responsible for the emergence of transport channels nearest to the Fermi level. These edge states, marked with orange and green curves, induce two-channel transport occurring strictly at the spatial edges of the structure, as depicted in Fig. \ref{transport}(a), corresponding to the spatially projected LDOS at the specific energies $E=0.37$ and $E=1.0$, respectively. These features highlight the potential for constructing such fractal devices and investigating the physics of the localized emergent states to develop novel device configurations.

\section{Conclusions}

Electronic and transport properties for topological Sierpinski Carpet systems within a $\pi$-flux tight-binding model were investigated. We have highlighted the importance of point mesh choices in determining the electronic properties and charge distribution within finite square lattices. Our analysis demonstrates that non-topological SCs preserve their main electronic features across varying mesh densities. This consistency is further supported also by the preservation of HOTI properties, even in the presence of disorder within such lattice systems. Nonetheless, our study reveals a sensitivity in the mesh partition to the spatial electronic distribution of neighboring states near the Fermi level, emphasizing the need for careful consideration in preserving fractal regularity and capturing the original physics underlying these fractal structures.

 In contrast to the conventional BBH model for the square lattice \cite{Benalcazar_2017, 2Benalcazar_2017}, our analysis identifies a transition occurring at a specific value of the $\gamma/\lambda$ ratio inside the quadrupolar phase. At this transition, the $E=0$ state can be exclusively populated by outer corner charges or by a combination of outer and inner anti-corner states, \textcolor{black}{helping the comprehension of the intricate charge dynamics within these systems.} The calculated spectral charges for corner and anti-corner states are consistent with the quadrupole moment at the HOTI system. Such spectral charges are fractional, underscoring their non-trivial topological nature. \textcolor{black}{To better characterize other phases of the presented  hopping parameter diagram, local real space marker \cite{PhysRevB.84.241106,PhysRevB.106.15512,KITAEV20062} should be used in analogy with the Chern number.}

We have also investigated the electronic transport properties in both dipolar and quadrupolar phases in SC chains. Our analysis identified two conductive channels, provided by localized states at the spatial edges of the system, in both regimes. This finding highlights the significance of these specific edge regions in the development of electronic devices exploring such characteristics within topological phases. The identification and understanding of these conductive channels gives valuable insights for smart and optimized designs  of electronic devices exploiting the  properties of the topological SCs. The overall picture of our findings contributes to a deeper understanding of the electronic behavior and the influence of lattice structure on such fractal systems.



\section*{Acknowledgments}
The authors would like to thank the INCT de Nanomateriais de Carbono for providing support on the computational infrastructure. LLL thanks the CNPq scholarship that provides condition for his research. AL would like to thank the FAPERJ grants, E-26/202.567/2019 and E-26/200.569/2023. The authors would like to express their gratitude to Tarik P. Cysne for valuable discussions.

\section*{Data Availability Statement}
The original contributions presented in the study are in the article; further inquiries can be directed to the corresponding author.

\appendix
\setcounter{figure}{0}

\section{Non-Topological Sierpinski Carpet Mesh}

Understanding the behavior of the compact localized states (CLS) in the systems requires a detailed examination of the mesh point density. 

\begin{figure}[!h]
    \centering
    \includegraphics[width=7cm]{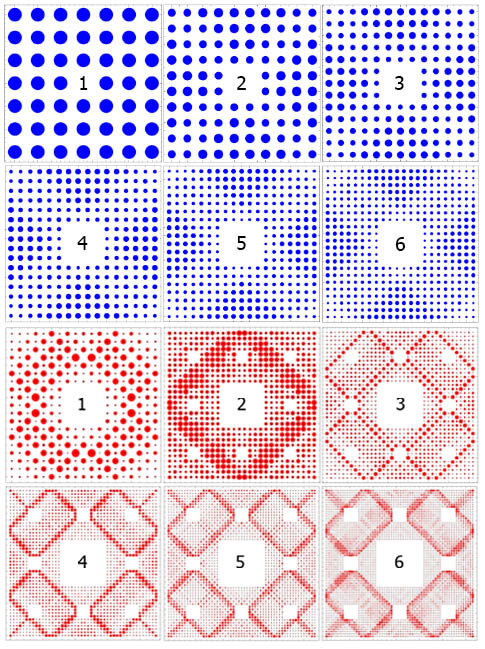}
   \caption{LDOS at $E=0$ for 6 different mesh points, numerated from 1 to 6, corresponding to generation 1 (blue panels) and 2 (red panels). All the LDOS are normalized by their maximum amplitude $|\psi_i|^2$.}
   \label{sup1}
\end{figure}

In Fig. \ref{sup1}, we compare the charge density at $E=0$ of the Sierpinski Carpet for six different mesh point densities shown in blue for G(1) and in red for G(2) SCs. The LDOS change substantially as the mesh point density increases. For larger meshes, however, the system LDOS converges to a saturated charge configuration, as may be concluded from the results labeled as 5 and 6 [see Fig. \ref{sup1}] for each one of the studied generations.

The number of sites, $S^{m}(n)$, and degeneracy $D^{m}[G(n)]$, other than $D[G^{m}(n)]$ of each one of the considered mesh point density systems are displayed in Table \ref{tab1}. The number of sites, is computed following the recursive relations,
\begin{gather}
    S^{m}(n)=8[S^{m}(n-1)-L^{m}(n-1)] \,\,,
    \\
    L^{m}(n)=3L^{m}(n-1)-2\,\,,
\end{gather}
with $S^{m}(0)$ and $L^{m}(0)$ given in Table \ref{tab1}.

\begin{table}[h]
    \centering
    \begin{tabular}{|c|c|c|c|c|c|c|}
        \hline
        Mesh m & $S^{m}(0)$ & $L^{m}$(0)  & $S^{m}(1)$&  $S^{m}(2)$ & $D^{m}[G(1)]$ & $D^{m}[G(2)]$ \\
        \hline
        2 & 16 & 4 & 96 & 688 & 8 & 16 \\
        \hline
        3 & 25 & 5 & 160 & 1176 & 10 & 24 \\
        \hline
        4 & 36 & 6 & 240 & 1792 & 12 & 32 \\
        \hline
        5 & 49 & 7 & 336 & 2536 & 14 & 40 \\
        \hline
        6 & 64 & 8 & 448 & 3408 & 16 & 48 \\
        \hline
        7 & 81 & 9 & 576 & 4408 & 18 & 56 \\
        \hline
    \end{tabular}
    \caption{Mesh point density table with the number of sites, $S^{m}(1)$ and $S^{m}(2)$, and the corresponding degeneracy $D^{m}[G(1)]$ and $D^{m}[G(2)]$ for the central state $E = 0$, for the first and second SC generations.}
    \label{tab1}
\end{table}

Depending on the choice of mesh points used to describe the SC, the degeneracy number of states at $E=0$ changes as linear monotonic functions, as shown in Table \ref{tab1} for the two first  SC generations. While for the first generation G(1) the increasing factor is two states from one mesh $m$ to the subsequent one, $m+1$, for the second generation 8 new degenerate states are added for increasing mesh partitions. 

\section{Considerations on Topological SC}

\begin{figure}[!h]
  \centering
    \includegraphics[width=8.3cm]{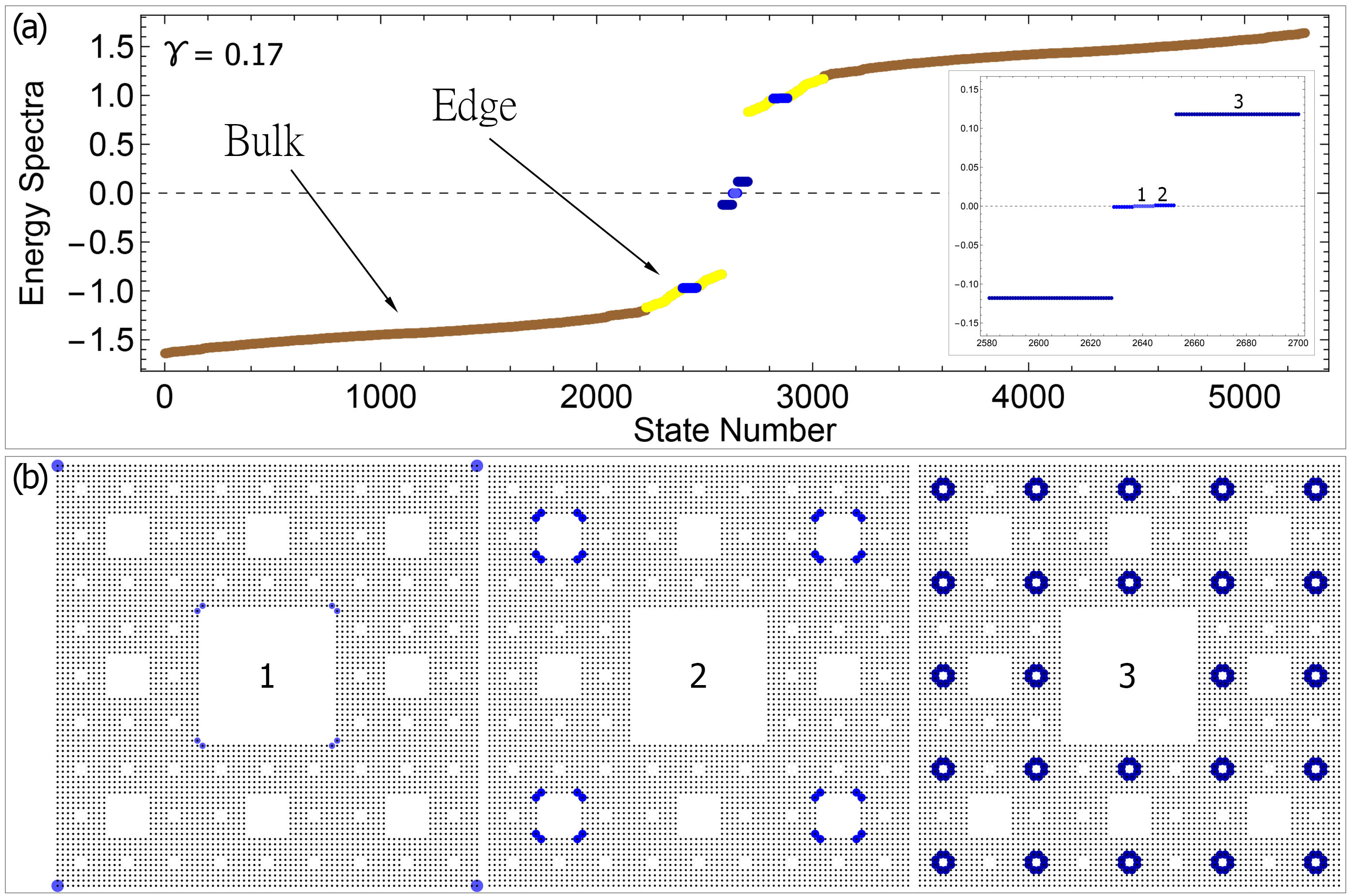}
   \caption{Topological Sierpinski Carpet G(3) [Mesh 2] with $\gamma_{x(y)}=0.17$ and $\lambda_{x(y)}=1.0$: (a) Energy spectra for quadrupolar phase and (b) LDOS for states 1, 2, and 3 near the Fermi level.}
   \label{sup_fig2}
\end{figure}

The energy spectra and LDOS calculations for a topological G(3) SC with $\gamma_{x(y)}=0.17$ and $\lambda_{x(y)}=1.0$ are shown in Fig. \ref{sup_fig2}(a). The results illustrate the relationship between fractality and self-similar charge distributions when compared to a G(2) SC system described in the main text. States near the Fermi level, labeled 1, 2, and 3 (blue symbols), are magnified in the inset, showing degeneracies of 8, 8, and 48, respectively. The corresponding LDOS for these states is presented in Fig. \ref{sup_fig2}(b) with different blue shades for states 1, 2, and 3. Notably, states 1 appear at the outer corners and internal anti-corner sites of the G(1)-fractal order, while \textcolor{black}{2-like states} are found at the anti-corner sites of four squares in the G(2)-fractal order.

\begin{figure}[!h]
    \centering
    \includegraphics[width=8.2cm]{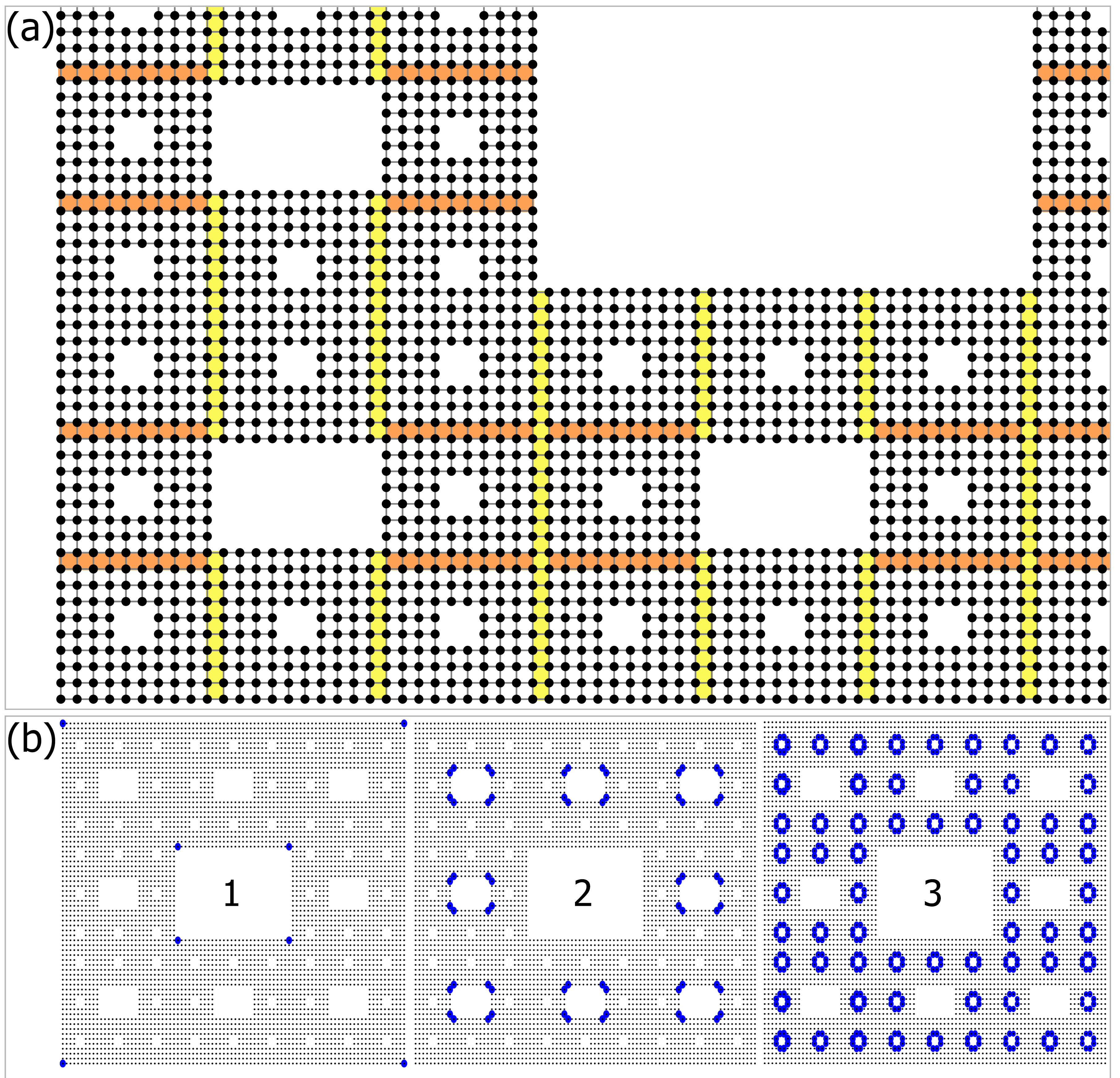}
   \caption{G(3) SC in the quadrupolar phase. (a) Non-symmetrical mesh illustration: orange and yellow shaded regions illustrate a symmetry break in the fractal grid. (b) LDOS for $\gamma_{x(y)}=0.17$ and $\lambda_{x(y)}=1.0$ for the three states nearest to the Fermi energy spectra (states 1, 2, and 3 of Fig. \ref{sup_fig2}).}
   \label{sup_fig4}
\end{figure}

Otherwise, the 3-like states are displayed in 24 of the 64 squares emerging at the G(3) order, creating a relatively regular pattern, similar to that shown in Fig. \ref{locator}(b) for a G(2)-SC. To explore the influence of this regular pattern on the partition sites in the $\pi$-flux of the Sierpinski Carpet, we introduce a break in the fractal symmetry of the square grid, indicated by orange and yellow strips in Fig. \ref{sup_fig4}(a). Other effects of disorder on the topological phases of Sierpinski lattices have been discussed in the context of spatial \cite{Zheng2022observation} and Anderson-type disorder \cite{Chin2024}.

The shaded ribbons represent intentional changes in the distances within the original mesh. We re-examine the LDOS for the three energies closest to the Fermi level, as shown previously in Fig. \ref{sup_fig2}. The results show that, with the introduced disorder, the 1-like state does not exhibit the previous symmetric anti-corner states; rather, internal corner states appear. Further, the charge distribution at the internal anti-corner sites now spatially fills all of the G(2) and G(3) squares for the state 2- and 3-like, respectively [central and right panels in Fig. \ref{sup_fig4}(b)]. An important point is that the fractional charge spectra are maintained for such corner-like states. Based on these findings, it is reasonable to conclude that small variations in the mesh might allow or prevent destructive quantum interference between the wave functions at special symmetric locations, which could impact the spatial localization of the charge density and can be externally engineered.

\bibliography{test}

\end{document}